# The last sunset on mainland Europe

*Jorge Mira Pérez*

(jorge.mira@usc.es)

Departamento de Física Aplicada
Universidade de Santiago de Compostela
E-15782 Santiago de Compostela, Spain

# *Abstract*

Certain coastal localities attempt to attract tourists by claiming to have the earliest sunrise or latest sunset in their country or even continent. This paper documents the places in mainland Europe where the sun sets latest, according to Coordinated Universal Time (UTC), at one or more days during the year. In contradiction to the naïve assumption that the sun always sets latest at the westernmost point, the point of last sunset changes cyclically over the course of a year due to the changing orientation of the axis of the Earth with respect to the sun. On mainland Europe, the focus of this study, between the winter and summer solstices the last sunset shifts successively from Cabo de São Vicente (Portugal) to Cabo da Roca (Portugal) to Cabo Touriñán (Spain) to a site near Aglapsvik (Norway) to a location in the Norwegian municipality of Måsøy south of Havøysund; and it shifts back again between the summer and winter solstices. There are even two days of the year (April 24th and August 18th) on which the last sunset of mainland Europe coincides with the last sunset of mainland Africa, at a site in Western Sahara near Cap Blanc. A similar analysis of the first sunrise in Spain is also done.

# 1. Introduction

To contemplate the sun going down in the sea is a delightful experience, especially when enjoyed at a site at which this event has some special geographical or historical significance—a circumstance often used to attract tourists. In Spain, the country with the third largest number of foreign visitors (UNWTO 2015), there are several such spots. For example, on the Costa da Morte, in Galicia (NW Spain), Cape Finisterre and the homonymous village (Fisterra in Galician) take their names from the Latin *Finis Terrae* - the end of the earth. According to the Roman historian Lucius Annaeus Florus (c. 74 CE – c. 130 CE), the name was coined by the Roman general Decimus Junius Brutus, who after conquering central and northern Portugal reached what was believed to be the very end of the world (Herrero 2009), and did not want to leave without seeing the sun sink in the sea (Murguía and Vicetto 1865; Romero Masia and Pose Mesura 1988, 94, quoting Florus, I, 33,12). There is some evidence supporting legends of pre-Roman pilgrimages to Finisterre to adore the sun, and that this may have been the origin of the Way of St. James, the Christian pilgrimage to the tomb of St. James 60 km away in Santiago de Compostela, which was established in the Middle Ages and persists to this day (Alonso Romero 2002; Herrero 2009; Murguía and Vicetto 1865; Romero Masia and Pose Mesura 1988, Sánchez-Carretero 2015).

Actually, Cape Finisterre is not the westernmost point of mainland Spain. That point is 20 km away to the north, at Cabo Touriñán (43º03′ N 9º18′ W), although the difference in longitude, less than one arc-minute, could not have been detectable by the Romans.

And the westernmost point of mainland Europe—the spot that would have been the true *finis terrae* for the Romans—is neither Cape Finisterre nor Cabo Touriñán, but 12 arc-minutes to the west at Cabo da Roca in Portugal (38º47′ N 9º30′ W). Nonetheless, every summer thousands of holidaymakers gather at Cape Finisterre and nearby locations to see the sun set, many of them in the belief that they are enjoying the last rays of sun to fall directly on mainland Europe that day. Are they wrong? The answer is "not always", at least if the approximately 4-second difference between the sunset times of Cape Finisterre and Cabo Touriñán is deemed negligible. This paper shows that, over the course of a year, the last sunset of mainland Europe oscillates among several coastal locations ranging from southern Portugal to above the Arctic Circle, and lists the dates on which it occurs at each of these locations.

An analogous situation holds in regard to sunrise. It is often claimed that Punta de s'Esperó in Menorca, as the easternmost point of Spain, is where each day's dawn may first be beheld in Spain. This supposed priority is exploited as a tourist attraction as witnessed by a webcam broadcasting the "first sun of Spain" on the nearby town of Es Castell's webpage. But how true is the claim? It turns out that it is true for about eight months of the year, and false for the rest.

Throughout this article, unless otherwise indicated, "time" refers to Coordinated Universal Time (UTC), not local solar times and local standard times (Spain, Portugal and Norway follow daylight saving time). The theoretical calculations of sunrise and sunset times presented apply to sea level only. Actual times when the sun is observed to rise and set at a particular location diverge from the theoretical calculation, due to atmospheric conditions, air pressure, atmospheric composition, and other variables such as the effects of atmospheric refraction on the time of sunset at higher latitudes, which create even greater discrepancies in the Arctic area. In addition, the dates on which the last sunset switches from one point to another, oscillate between one day and the next with a period of a few years (mainly due to the leap year calendar), as do those of the equinoxes and solstices.

Coastlines provided by Natural Earth (http://www.naturalearthdata.com) were plotted by gnuplot (http://www.gnuplot.info) along with the border between the illuminated and dark side of the Earth (the *terminator*). Then, Xplanet (http://xplanet.sourceforge.net) was used to generate the orthographic projections and labels shown in the illustrations.

## 2. Basic considerations

Given two locations on Earth, A and B, with A south of B at exactly the same longitude, then high noon—defined as the time at which the sun is due south in the northern hemisphere and due north in the southern—always occurs at the same time at A as at B. But the sun sets at the same time at A as at B only on two days of the year, the equinoxes, when the Earth's axis of rotation is tilted neither away from nor towards the sun, and the terminator accordingly coincides with a meridian (see Figure 1a; we ignore twilight zones due to refraction of light by the atmosphere). During the rest of the year, when the northern pole of Earth's axis is tilted away from or towards the sun, the sunset half of the terminator crosses meridians at an angle that depends on latitude. On a cylindrical map projection in which lines of latitude are horizontal and meridians are parallel and run north-south (such as a Mercator or Gall projection), in the northern hemisphere, looking north, it bends towards the east from the March equinox (around March 20[th]) to the September equinox (around September 23[rd]) and to the west from the September-to-March equinox, and *vice versa* in the southern hemisphere, looking south. Accordingly, in the northern hemisphere (the focus of this paper), between the spring and autumn equinoxes the sun sets later at B than at A (Figure 1b), and earlier between the autumn and spring equinoxes. Then too, as is well known, between the spring and autumn equinoxes the sun does not set at all in a zone above the Arctic Circle. As the terminator leans farther from the meridians, this zone creeps south (reaching mainland Europe on May 11[th] at Cape Nordkinn, Norway; see Figure 2) until at the summer solstice it begins to progress north again. As the terminator edges towards the meridians, the zone of perpetual daylight shrinks to nothing between the summer solstice and the autumn equinox. Then, between the autumn and spring equinoxes, a zone of perpetual night echoes this behaviour.

If location A lies west of location B and both have the same latitude, the sun always sets later at A than at B. But if A lies to southwest of B, the arcing of the terminator described above can cause the sun to set later at B than at A at some date between the spring and autumn equinoxes. Whether and when this occurs depends on the differences in their latitudes and longitudes.

## 3. Application to mainland Europe

There are many simulators available that map the terminator. Working with a temporal precision of one day, they suffice to show that between the winter and summer solstices the point of last sunset on mainland Europe shifts from Cabo de São Vicente (Portugal) to Cabo da Roca (Portugal), to Cabo Touriñán (Spain), and then to the complex coast of northern Norway. The coast of the French department of Finistère, the name of which suggests it might enjoy this distinction, always lies to the southeast of the last-sunset terminator and therefore is never the site of the last sunset on mainland Europe.

Because of the multitude of islands fringing the Norwegian mainland, finding the Norwegian last-sunset points requires inspection of the terminator at a relatively large scale and relatively high temporal precision, for which purpose the resources of the US Naval Observatory's Naval Oceanography Portal (http://www.usno.navy.mil/USNO/astronomical-applications/data-services/data-services) comes in handy. Supplying geographical coordinates with a precision of one arc-minute generates sunset time tables (UTC) with a precision of one chronological minute, which suffices for present purposes. For example, at a latitude of 45º, one minute of arc is equivalent to about 1.3 km, which is transited by the apparent sun in about 4 seconds. We ignore variation due to the elevation of the sunset observation point (the higher the elevation, the farther away the horizon will be and thus sunset will come later).

Close scrutiny shows that the first last-sunset point after Cabo Touriñán is at 69º28'N, 18º09'E on the Norwegian coast near Aglapsvik, a village on the mainland near the city of Tromsø. A few days later, the last sunset moves on again to a site with approximate coordinates 71º00'N, 24º39'E in the municipality of Måsøy, south of Havøysund (Figure 3). Neither of these Norwegian last-sunset points is either the westernmost or the northernmost point of mainland Norway, which are respectively Vardetangen (60º48'N 5º56'E) and Cape Nordkinn (71º08'N 27º39'E).

In successive years, the dates on which the last sunset switches from one point to another, like those of the equinoxes and solstices, may change by a day with a period of a few years. Also, there are also periods in which, to meaningful precision, two points share the last sunset. With these provisos, the periods in which the last sunset occurs at each point are listed in Table 1 together with the UTC time of sunset at the transition dates. It is striking that within three or four days after the spring equinox the bending of the terminator is already sufficient to shift the point of last sunset from Cabo da Roca to Cabo Touriñán. The shift from Cabo Touriñán to the Aglapsvik area (Figure 4) takes place on April 24$^{th}$ (at these latitudes and time of year, the rapid change in day length facilitates identification of a single transition date, in contrast to the more gradual transitions from Cabo de São Vicente to Cabo da Roca or from Cabo da Roca to Cabo Touriñán). Then, on May 1$^{st}$ the last sunset moves on to Måsøy (Figure 3).

On May 11$^{th}$, the Måsøy site is absorbed by the waxing zone of midnight sun (Figure 2). The edge of this zone is the northernmost border of the area in which sunsets occur, and its westernmost point on the mainland, which during this period may be regarded as the point of last sunset, moves southward along the coast from day to day until it reaches the Arctic Circle at the summer solstice. Technically, then, on the days the northern European mainland lies within the midnight sun zone, there is no last European sunset.

After the summer solstice, the same sequence of events occurs in reverse, at dates symmetrical with respect to the solstice, as the terminator gradually unbends. First the border of the midnight sun zone retraces its steps northwards, reaching the Måsøy site on August 1st. Then the last sunset shifts from Måsøy to Aglapsvik on August 11th, to Cabo Touriñán on August 18th, and to Cabo da Roca 3-4 days before the autumn equinox, after which the westward bending of the terminator leads, around October 19th (Figure 5), to a shift to Cabo de São Vicente (37º01'N 9º01'W), where the last sunset remains until around February 21st. Cabo de São Vicente thus enjoys the last sunset for 4 months, longer than any other point of mainland Europe.

Note that the times of sunset listed in Table 1 are not symmetric with respect to the summer solstice (although the actual dates of the transitions from one site to another are roughly symmetrical about the summer solstice). This reflects the difference between apparent solar time and UTC, which leads, for example, to locations in the northern hemisphere having their earliest sunset a few days before the winter solstice and their latest sunset a few days after the summer solstice.

Curiously, on April 24th and August 18th, at the time of the shift from Cabo Touriñán to Aglapsvik or *vice versa*, the last sunset of mainland Europe coincides with that of Land's End (the westernmost point of mainland England) and with the last sunset of mainland Africa (Figure 6), which takes place in Western Sahara near Ras Nouadhibou (Cap Blanc, approximate coordinates 20º50'N 17º06'W). Once more, it is significant that Cap Blanc is not the westernmost point of mainland Africa, which is Cap Vert in Senegal.

## 4. The first sunrise of Spain

While half of the terminator marks points at which the sun is setting, on the half arcing across the opposite hemisphere the sun is rising. Certain localities, such as the town of Es Castell in Menorca (http://www.aj-escastell.org/), claim to receive the first sunrise for some territory—in this case Spain. Consistent with this claim, the town has a monument to Eos, the Greek goddess of the dawn, and has installed a webcam to broadcast its sunrise on its official web page. Certainly, Punta de s'Esperó (39º53'N 4º20'E), which is near Es Castell though actually located in the municipality of Maó, enjoys the first sunrise in Spain for most of the year, approximately from around August 21$^{st}$ to April 21$^{st}$. But during the rest of the year the westward bending of the terminator leads to the sun first rising on Spain at the easternmost point of the Spanish mainland, Cap de Creus (42º19'N 3º19'E).

## 5. Conclusions

Given the spatial and temporal precision of this study, explained above, on any given day between August 1$^{st}$ and May 11$^{th}$ the last sunset in mainland Europe observable at sea level takes place at one of five points:

- Cabo de São Vicente, Portugal (for around 4 months, approximately from October 19$^{th}$ to February 21$^{st}$)

- Cabo da Roca, Portugal (the westernmost point of mainland Europe, between February 21$^{st}$ and a few days after the spring equinox, and from a few days before the autumn equinox until October 19$^{th}$)
- Cabo Touriñán, Spain (approximately from March 24$^{th}$ to April 24$^{th}$ and from August 18$^{th}$ to September 19$^{th}$)
- A site near Aglapsvik on the mainland south of Tromsø, Norway (approximately from April 24$^{th}$ to May 1$^{st}$ and from August 11$^{th}$ to August 18$^{th}$)
- A site in the municipality of Måsøy south of Havøysund, Norway (approximately from May 1$^{st}$ to May 11$^{th}$ and from August 1$^{st}$ to August 11$^{th}$).

Between May 12$^{th}$ and August 1$^{st}$ northern mainland Europe experiences the midnight sun, and during this period, strictly speaking, either there is no last sunset on mainland Europe or else it occurs at the intersection of the Norwegian coastline with the edge of this zone, which moves south until it reaches the Arctic Circle at the summer solstice and then returns north. During the switches between Cabo Touriñán and Aglapsvik (around April 24$^{th}$ and August 18$^{th}$), the last sunset of mainland Europe coincides with the last sunset of mainland Africa, which occurs in the Western Sahara near Ras Nouadhibou (Cap Blanc). The first sunrise of Spain takes place at Punta de s'Esperó in Menorca during approximately eight months (August 21$^{st}$ to April 21$^{st}$), and at Cap de Creus in Catalonia for the rest of the year. It will be for others to apply these observations to their efforts to attract tourists to deserving localities.

# Acknowledgements

The author is indebted to José María Martín Olalla, from the Departamento de Física de la Materia Condensada of the Universidad de Sevilla (Spain), for his help with the figures of this work.

# References

Alonso Romero, F. 2002. *Historia, leyendas y creencias de Finisterre* [History, legends and beliefs of Finisterre], Briga Edicións.

Herrero N. 2009. "La atracción turística de un espacio mítico: peregrinación al cabo de Finisterre" [The touristic attraction of a mythical space: pilgrimage to Cape Finisterre] *Pas*os 7 (2): 163-178

Murguía, M., and B. Vicetto. 1865. *Historia de Galicia* [History of Galicia] [Ed. La Gran Enciclopedia Vasca (1980) – vol. I].

Romero Masia, A. M. and X. M. Pose Mesura. 1988. *Galicia nos textos clásicos* [Galicia in classical texts] Monografías do Museo Arqueolóxico da Coruña.

Sánchez-Carretero, C. (ed.). 2015. *Heritage, Pilgrimage and the Camino to Finisterre. Walking to the End of the World*, GeoJournal library 117, Springer.

United Nations World Tourism Organization (UNWTO). Tourism Highlights. 2015. http://www.e-unwto.org/doi/pdf/10.18111/9789284416899 (see page 6).

*Table 1. Places and times of year at which the last sunset of mainland Europe can be observed . The dates are indicated with a precision of ±1 day. For each place and date, the UTC sunset time is indicated with a precision of ±1 minute. Twilight effects are not taken into account, and neither is the elevation of the observation point.*

| *LOCATION* | *DATES* | *SUNSET TIME* |
|---|---|---|
| **Cabo de São Vicente** *37°01'N 9°00'W* | October 19 | 17:54 |
| | February 21 | 18:22 |
| **Cabo da Roca** *38°47'N 9°30'W* | February 21 | 18:22 |
| | March 24 | 18:54 |
| **Cabo Touriñán** *43°03'N 9°18'W* | March 24 | 18:54 |
| | April 23 | 19:29 |
| **Aglapsvik** *69°28'N 18°09'E* | April 24 | 19:31 |
| | May 1 | 20:06 |
| **Måsøy** *71°00'N 24°39'W* | May 1 | 20:06 |
| | May 10 | 21:29 |
| **Norwegian coast*** | May 11 | -- |
| | August 1 | -- |
| **Måsøy** *71°00'N 24°39'W* | August 2 | 21:29 |
| | August 10 | 20:18 |
| **Aglapsvik** *69°28'N 18°09'E* | August 11 | 20:12 |
| | August 18 | 19:35 |
| **Cabo Touriñán** *43°03'N 9°18'W* | August 18 | 19:35 |
| | September 19 | 18:40 |
| **Cabo da Roca** *38°47'N 9°30'W* | September 20 | 18:39 |
| | October 19 | 17:54 |

[*] During this period the westernmost point of the edge of the midnight sun zone in mainland Europe skips along the coast from Måsøy to the Arctic Circle and back.

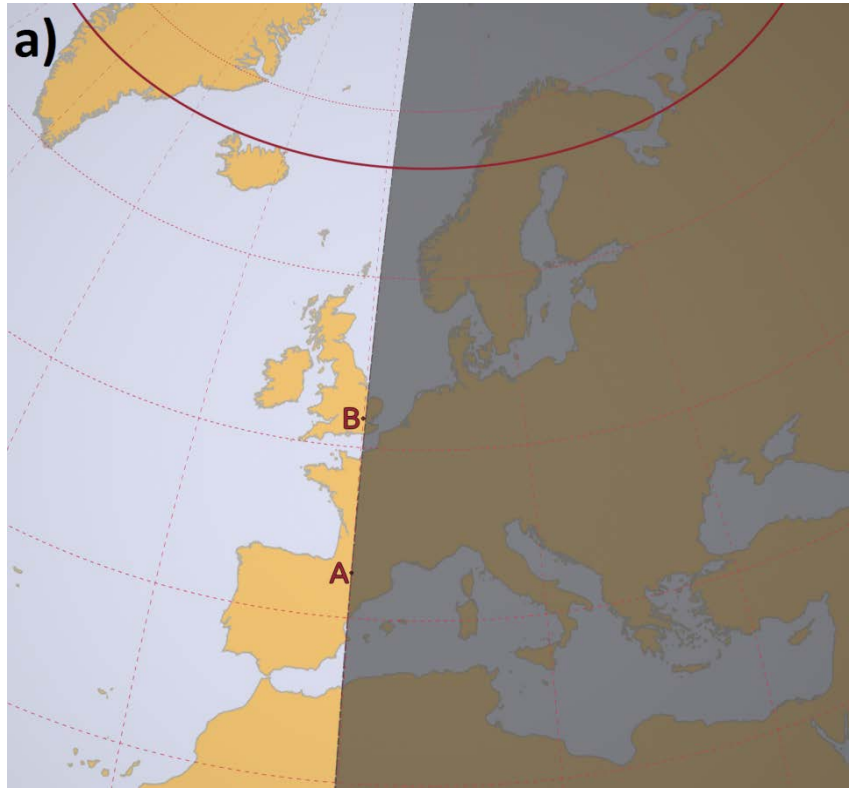

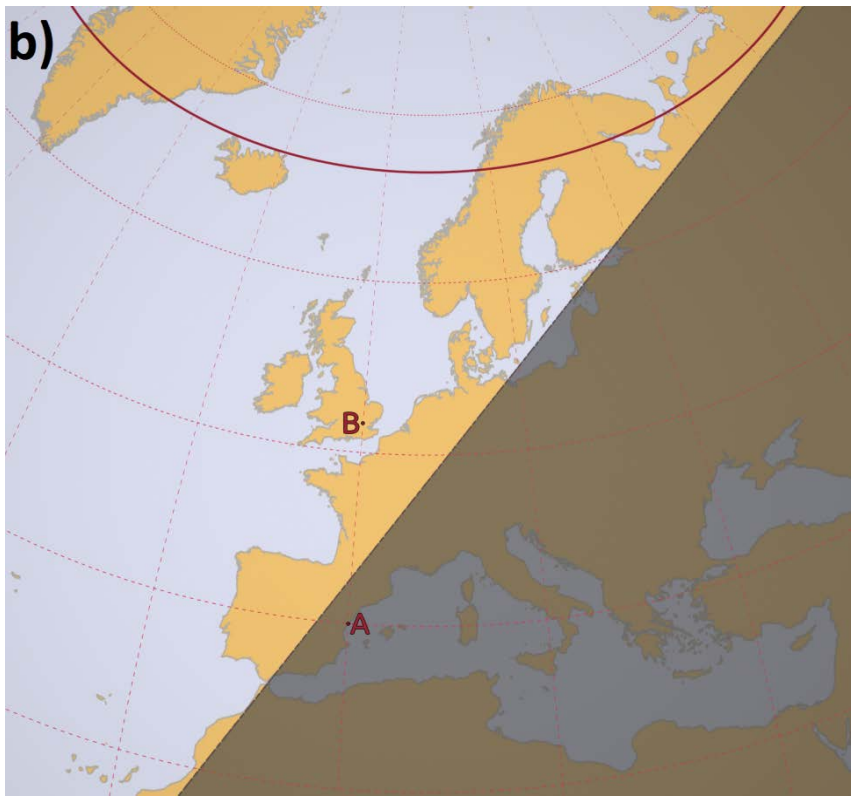

Figure 1. a) At the equinoxes, sunset occurs at the same time at points A and B lying on the same meridian. b) Between the spring and autumn equinoxes the sun sets later at the more northerly point, B.

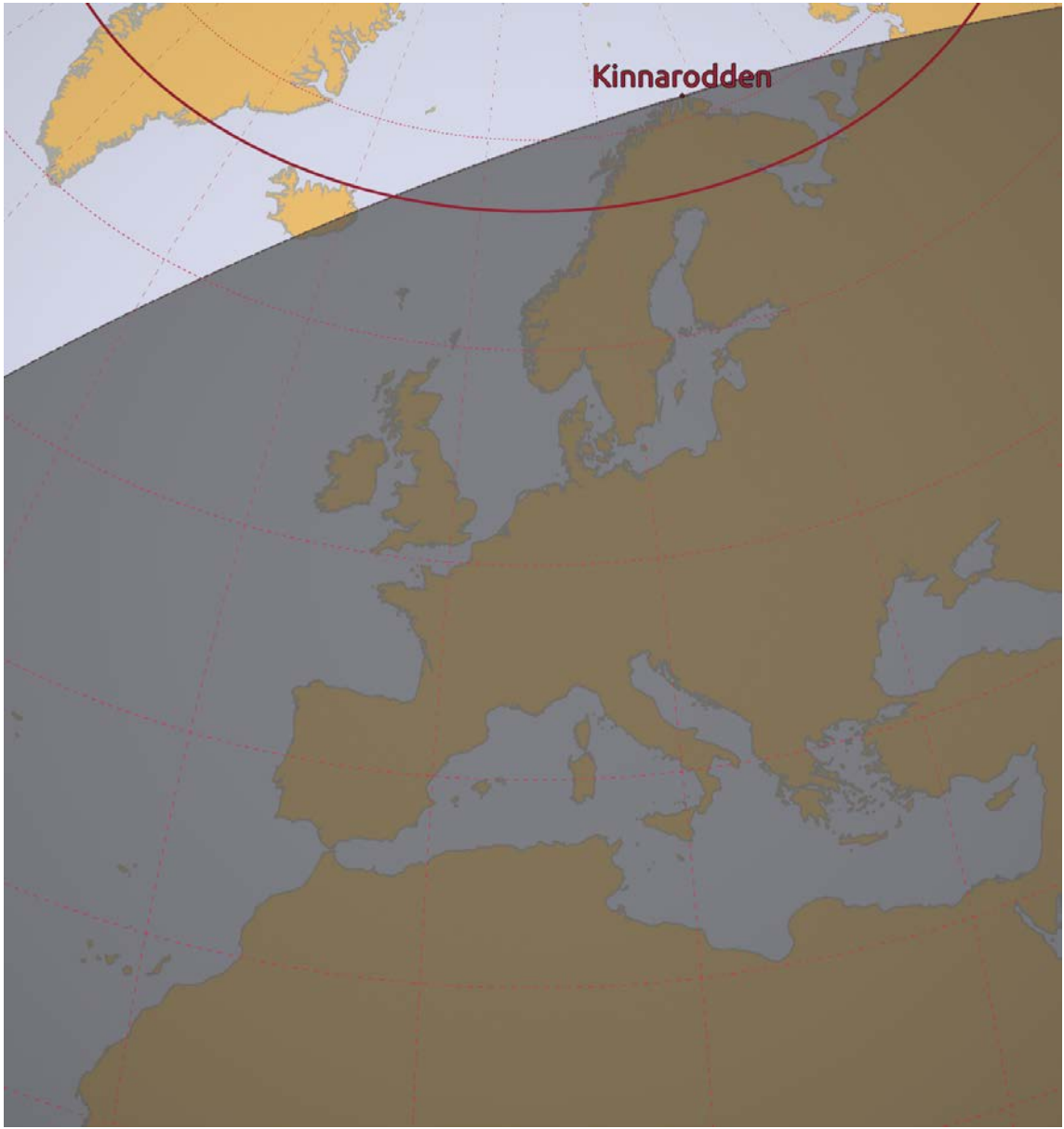

Figure 2. The zone of midnight sun reaches mainland Europe at Cape Nordkinn (Kinnarodden) on May 11th and recedes therefrom on August 1st.

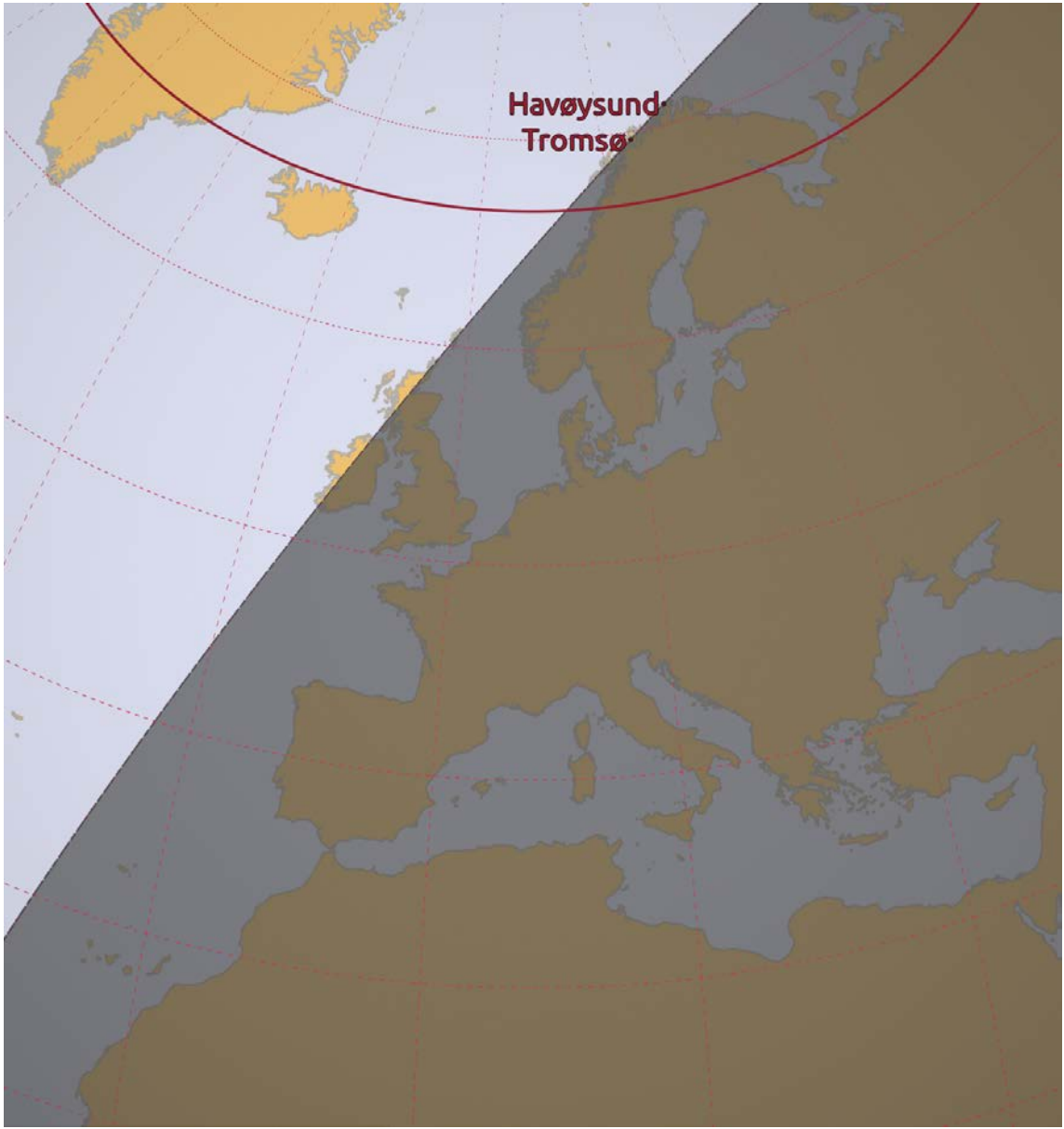

Figure 3. Position of the terminator when the last sunset times are similar in Aglapsvik (near Tromsø) and Måsøy (south of Havøysund), on May 1st and August 11th.

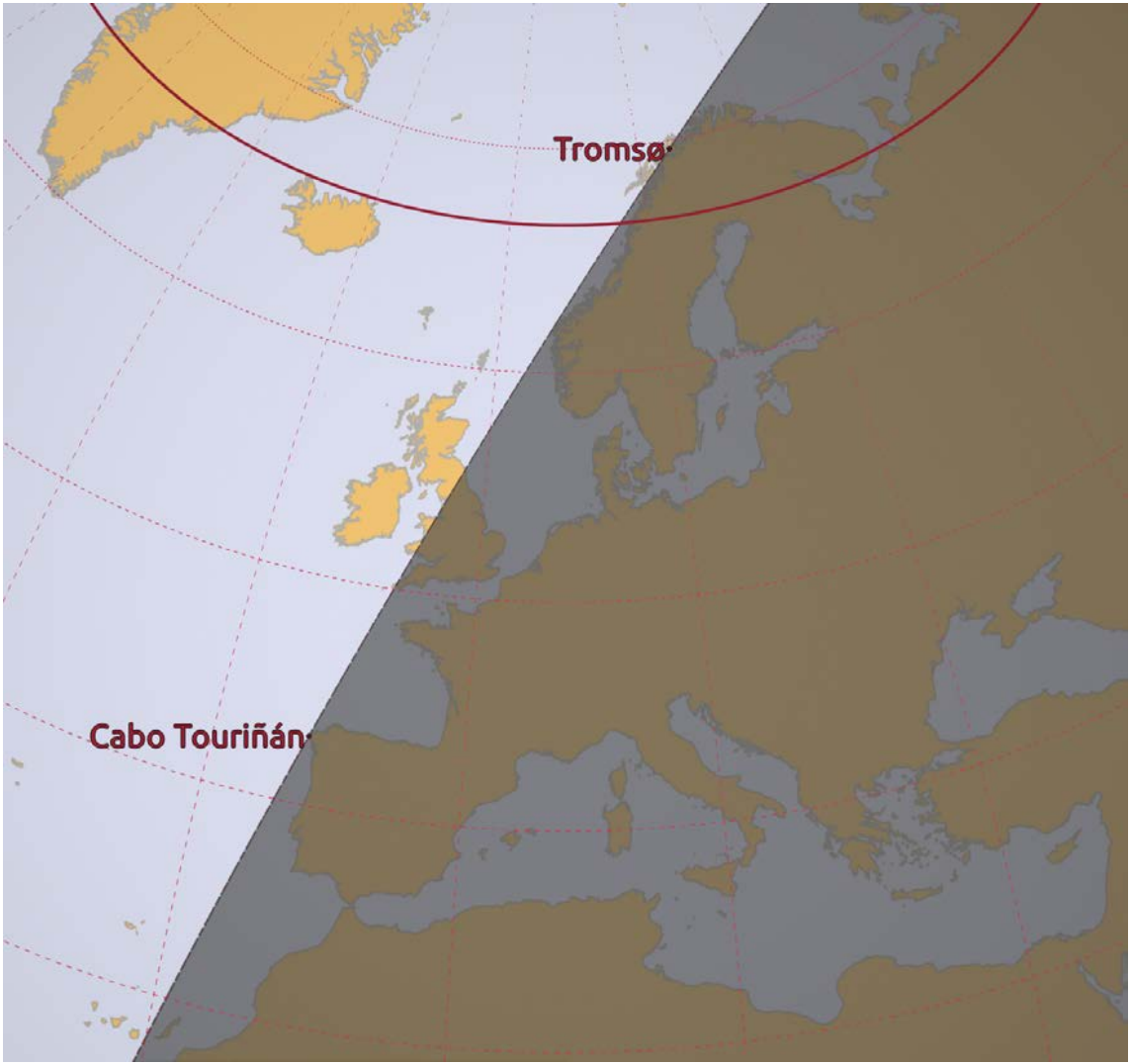

Figure 4. The last sunset shifts between Aglapsvik and Cabo Touriñán on April 24th and August 18th.

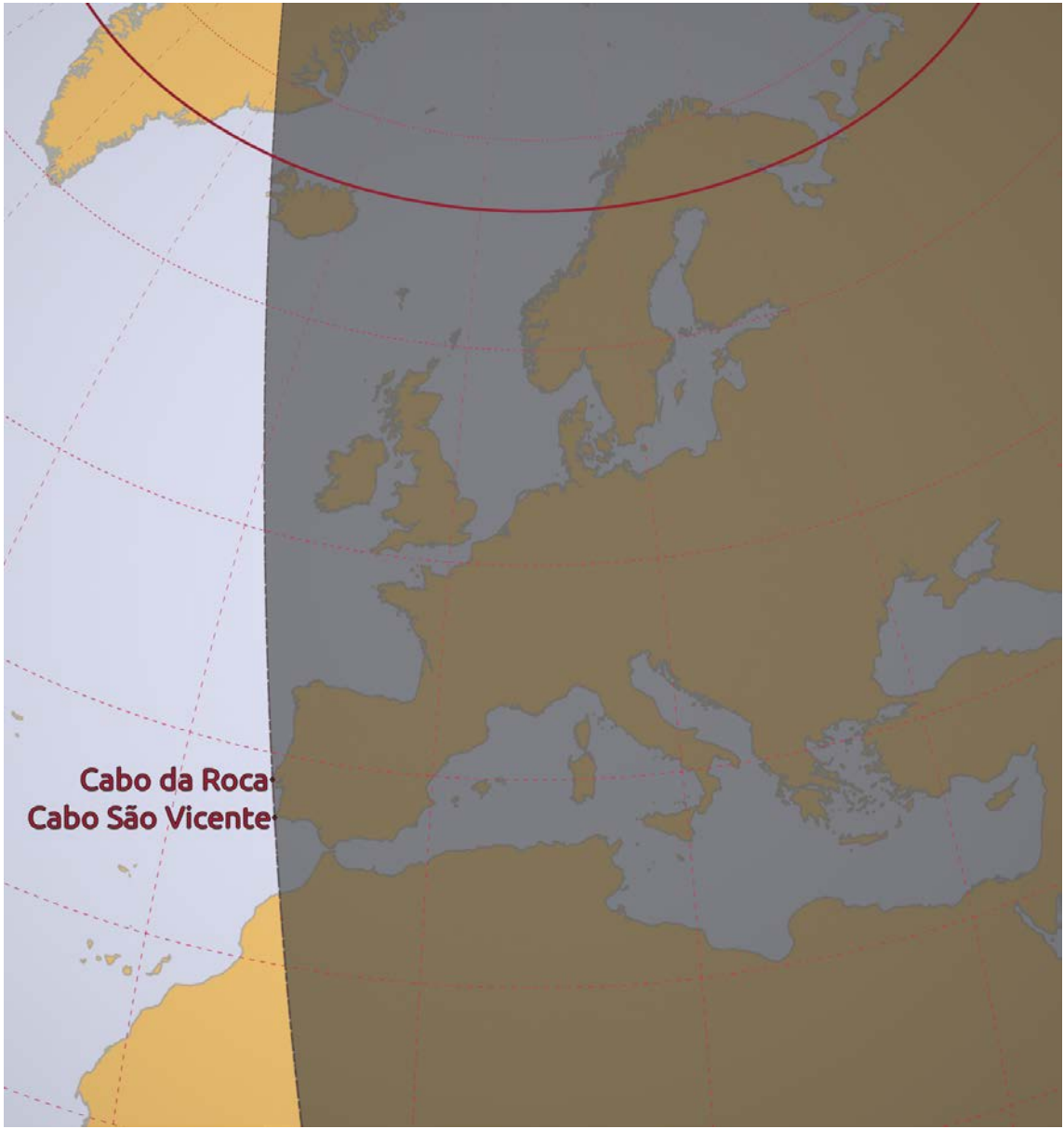

Figure 5. After the autumn equinox the terminator bends to the west, which results in the last sunset shifting from Cabo da Roca to Cabo de São Vicente on October 19th. The reverse shift occurs on February 21st.

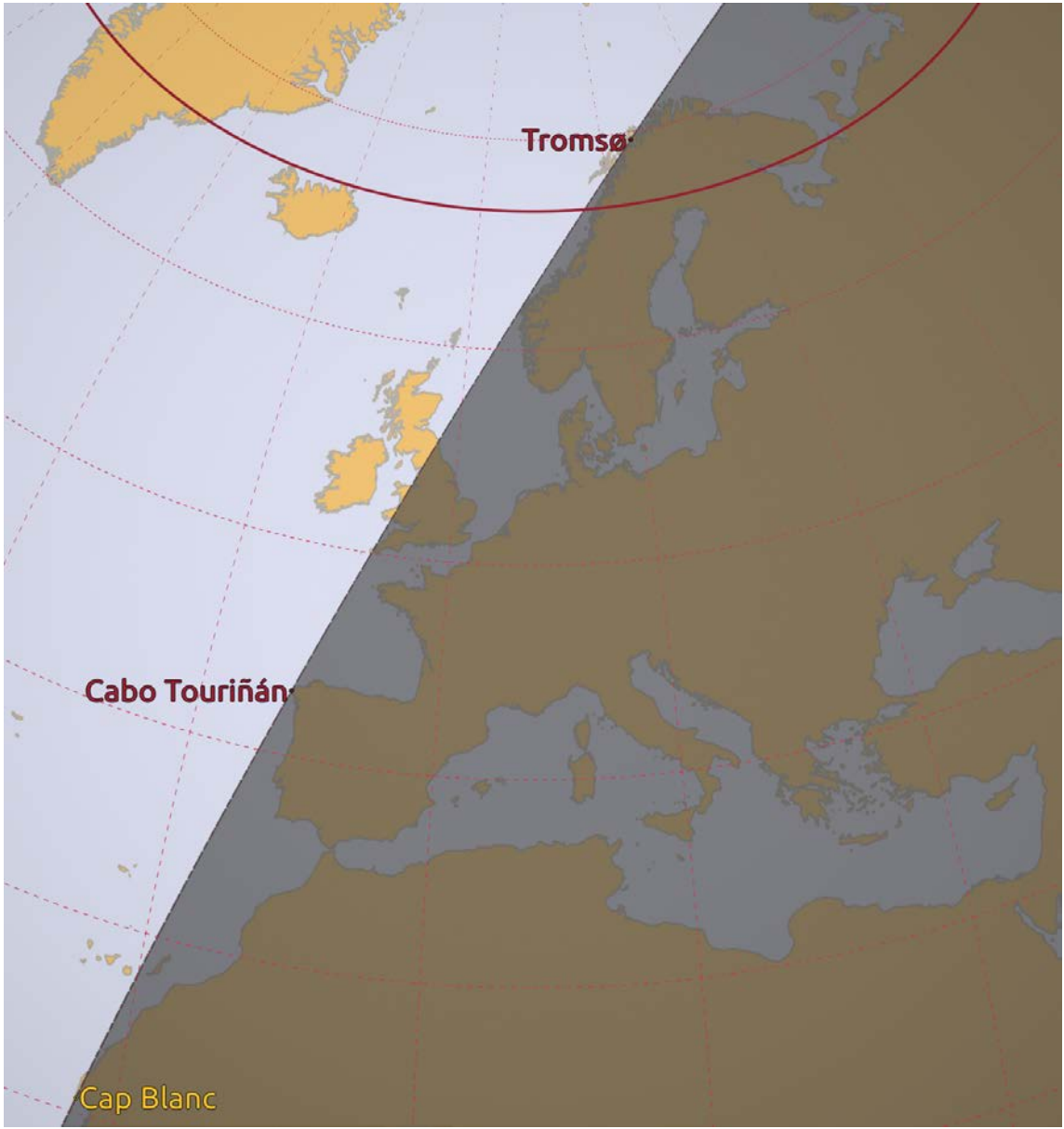

Figure 6. During the shift between Cabo Touriñán and Aglapsvik (April 24th and August 18th), the last sunset of mainland Europe coincides with that of mainland Africa (and also with that of Land's End, the westernmost point of mainland England).